# The Influence of Commercial Intent of Search Results on Their Perceived Relevance

Dirk Lewandowski

Hamburg University of Applied Sciences
Department Information
Finkenau 35
D – 22081 Hamburg
Germany

dirk.lewandowski@haw-hamburg.de

## ABSTRACT
We carried out a retrieval effectiveness test on the three major web search engines (i.e., Google, Microsoft and Yahoo). In addition to relevance judgments, we classified the results according to their commercial intent and whether or not they carried any advertising. We found that all search engines provide a large number of results with a commercial intent. Google provides significantly more commercial results than the other search engines do. However, the commercial intent of a result did not influence jurors in their relevance judgments.

## Categories and Subject Descriptors
H.3.3 [**Information Search&Retrieval**]: Retrieval models, Search process, Selection process; H.3.5. [**On-line Information Services**]: Web-based services

## 1. INTRODUCTION
Search engines are still of great interest to the information science community. Numerous tests have been conducted, focusing especially on the relevance of the results provided by different web search engines. In this paper, we extended the research body on retrieval effectiveness tests through additionally classifying the results according to their commerciality—that is, according to the commercial intent of a web page (e.g., selling a product) and the presence of advertising on the page.

We can assume that search engine results are somehow biased towards commercial results when we consider that there is a whole industry focusing on improving certain websites' ranking in the major search engines. This so-called "search engine optimization" (SEO) may have an effect on the results rankings in general. However, it is also probable that search engines counteract this by favoring non-commercial results. For a better understanding of search engines not only as tools for information retrieval, but also as tools that have an impact on knowledge acquisition in society, it is important to identify any flaws in their results presentation.

While some studies have considered *users* having a commercial intent (i.e., users searching for e-commerce offerings) and have asked whether advertising results are judged comparably to organic results [19-21], to our knowledge no studies have yet taken into account to what degree search engines display results with a commercial intent and how these results are judged in comparison to non-commercial results.

The rest of this paper is structured as follows: first, we give an overview of the literature on the design and the results of search engine retrieval effectiveness tests. Then, we present our research questions, followed by our methods and test design. After that, we present our findings. We close with conclusions, including suggestions for further research.

## 2. LITERATURE REVIEW
In the following review, we focus on studies using English- or German-language queries. English-language studies make the most part of studies conducted, while we added the German studies because we used German queries in our investigations, as well. However, we fortunately saw that researchers from other countries are increasingly interested in web search, which led to a considerable body of research on searching in other languages than English (for a good overview, see [23]). In our review, only studies using informational queries (cf. [2]) will be considered. While there are some studies on the performance of web search engines on navigational queries (see [13, 27], these should be considered separately because of the very different information needs considered. However, there are some "mixed studies" (e.g., [11]) that use query sets consisting of informational as well as navigational queries. These studies are flawed because in contrast to informational queries, the expected result for a navigational query is just one result. When more results are considered, even a search engine that found the desired page and placed it on the first position would receive bad precision values. Therefore, informational and navigational queries should be strictly separated.

In general, retrieval effectiveness tests rely on methods derived from the design of IR tests [31] and on advice that focuses more on web search engines [9, 14]. However, there are some remarkable differences in study design, which are reviewed in the following paragraphs.

First, the number of queries used in the studies varies greatly. Especially the oldest studies [4, 7, 24] use only a few queries (5 to 15) and are, therefore, of limited use (for a discussion of the minimum number of queries that should be used in such tests, see [3]). Newer studies use at least 25 queries, some 50 or more.

In older studies, queries are usually taken from reference questions or commercial online systems, while newer studies focus more on the general users' interest or mix both types of questions. There are studies that deal with a special set of query topics (e.g., business; see [9]), but there is a trend in focusing on the general user in search engine testing (cf. [26]).

Regarding the number of results taken into account, most investigations only consider the first 10 or 20 results. This has to do with the general behavior of search engine users. These users only seldom view more than the first results page. While these results pages typically consist of 10 results (the infamous "ten blue links"), search engines now tend to add additional results (e.g., from news or video databases), so that the total amount of results especially on the first results page is often considerably higher [18].

Furthermore, researchers found that general web search engine users heavily focus on the first few results [6, 10, 16, 22], which are shown in the so-called "visible area" [18] — i.e., the part of the first search engine results page (SERP) that is visible without scrolling down. Keeping these findings in mind, a cut-off value of 10 in search engine retrieval effectiveness tests seems reasonable.

An important question is how the results should be judged. Most studies use relevance scales (with three to six points). Griesbaum's studies [11, 12] use binary relevance judgments with one exception: results can also be judged as "pointing to a relevant document" (i.e., the page itself is not relevant but has a hyperlink to a relevant page). This is done to take into account the special nature of the web. However, it seems problematic to judge these pages as (somehow) relevant, as pages could have many links, and a user then (in a worst case scenario) has to follow a number of links to access the relevant document.

Two important points in the evaluation of search engine results are whether the source of results is made anonymous (i.e., the jurors do not know which search engine delivered a certain result) and whether the results lists are randomized. The randomization is important to avoid learning effects. While most of the newer studies do make the results anonymous (as to which search engine produced the results), we found only three studies that randomize the results lists [9, 26, 34].

Most studies use students as jurors. This comes as no surprise, as most researchers teach courses where they have access to students that can serve as jurors. In some cases (again, mainly older studies), the researchers themselves judge the documents [4, 7, 8, 11, 12, 24].

To our knowledge, there is no search engine retrieval effectiveness study that uses multiple jurors. This could lead to flawed results because different jurors may not agree on the relevance of individual results.

An overview of the major web search engine retrieval effectiveness tests conducted from 1996 to 2007 can be found in [26]. For an overview of older tests, see [9].

There are some studies dealing with search engines' bias in terms of representation of the web's content in their indices [32, 33]. Further studies deal with the general assumptions underlying the algorithms of results ranking [5]. However, to our knowledge no studies have dealt with the *intent* of search results—that is, the intentions of the content provider behind each result.

## 3. RESEARCH QUESTIONS
We conducted this study to address several research questions, as follows:

RQ1: How effective are web search engines in retrieving relevant results? (Results to this more general research question can be seen as a kind of by-product of our study.)

RQ2: What ratio of the top 10 search engine results are pages with a commercial intent?

RQ3: Are results with a commercial intent comparably relevant to non-commercial results?

RQ4: Do banner or textual ads have a negative effect on the perceived relevance?

## 4. METHODS
We describe our methods in two parts: first, we give a detailed overview of our retrieval effectiveness test, and then we explain our results classification.

### 4.1 Retrieval Effectiveness Test Design
For our investigation, we designed a standard retrieval effectiveness test using 50 queries. The test design was based on the work of Lewandowski [26]. The queries were sent to the search engines selected, and results were then made anonymous and were randomized. Each juror was given the complete results set for one query and judged the results as relevant or not relevant. Then, the collected results were again assigned to the search engines and results positions; results from the classification tasks were added. In the following paragraphs, we describe our test design in detail.

*4.1.1 Selection of Queries*
The queries used in this study were taken from a survey where students were asked for their last informational query posed to a web search engine. We asked for queries used in their private lives, not used for research for their studies. All queries were German language.

"Informational query" was defined, as in Broder's paper [2], as a query where the intent is to retrieve multiple pages and use them for getting information on a topic. One could also say that this kind of query should produce results that are suitable for a user to build his own opinion on a topic [25]. While the underlying intention of an informational query may be commercial, we assume that this is mostly not the case. The queries used in this study do not suggest a commercial intent. However, we did not explicitly ask for non-commercial queries.

Each student was asked to not only record the query itself, but also a short description of the underlying information need. This was used to help the jurors determine the intent of the user, and also to help them to judge the documents accordingly. To give an example, see the following query with description (translated into English):

> Query: cost of living japan
>
> Description: Seeking information on prices in Japan.

When comparing queries with descriptions, one can find that for some queries, the description describes a narrower information need than the query would suggest. However, this is a typical problem in web search, where users often assume that additional information to the query should be determined by the search engine.

We hoped to get queries that represent general user behavior, at least to some degree. Table 1 shows the distribution of query length for the queries used in this study. The average length of the queries used is 2.06 terms, while in general log file-based investigations, the average query length for German-language queries lies between 1.6 and 1.8 terms [17]. (German language queries are in general shorter than English-language queries due to the heavy use of compound words in German.) However, we think that the differences are not so great that they would influence the test significantly.

**Table 1. Distribution of query lengths.**

| Length | Number of queries | Percentage |
|---|---|---|
| One word | 12 | 24 |
| Two words | 25 | 50 |
| Three words | 11 | 22 |
| Four words | 2 | 4 |

### 4.1.2 Choice of Search Engines

For our investigation, we used the three major web search engines, namely, Google, Yahoo and Microsoft.[1] Our first criterion was that each engine should provide its own index of web pages. Many popular search portals (such as AOL) make use of results from the large search providers. For this reason, these portals were excluded from our study. The second criterion was the popularity of the search services. According to Search Engine Watch, a search industry news site, the major international search engines are Google, Yahoo, Microsoft's Bing, and Ask.com [30]. However, Ask.com does not play any significant role on the German search engine market [35], so we excluded this search engine from our study.

### 4.1.3 Number of Results

For each query, we analyzed the first 10 results from each search engine. A cut-off value had to be used because of the very large results sets produced by the search engines. As stated in the literature review section, studies found that users only look at the first few results and ignore the rest of the results list.

### 4.1.4 Relevance Judgments

In this study, we used binary results judgments. We did not use any particular definition of relevance or give the jurors any criteria to help judge whether a document was relevant or not. We are well aware of the ongoing discussion of how relevance should be defined (cf. [1, 15, 28, 29]), but we decided to let our jurors determine what is relevant and what is not. Due to the design of our test (it was only possible to give jurors written instructions), it also would have been very complicated to make jurors familiar with a certain concept of relevance.

### 4.1.5 Jurors

Undergraduate students from the author's institution were used as jurors. Each juror evaluated the results of one query.

The student jurors were different from the ones from whom the search queries were collected. Therefore, jurors had to rely on the descriptions of the query intents. We are aware that this is only the second-best solution, but due to the laborious work of collecting and preparing the results, we had to make some compromises.

### 4.1.6 Data Collection

As the search results were collected all at once and were printed out for further judgment, we omitted the problem of continuously changing results sets. Data was collected in summer, 2009.

Note that not all jurors judged all documents they were supposed to judge (i.e., some jurors omitted some documents they were given). Therefore, we did not reach the maximum number of 1,500 results (50 queries x 10 results x 3 search engines results), but only 1,402 results.

## 4.2 Results Classification

Commercial results in the sense used in this paper are results with a commercial intent as measured by who the provider of the individual result is. That is, a result from a university website is considered non-commercial because the provider (the university) is a non-commercial organization. Classification was done by a student research assistant at the author's institution. As the classification tasks were unambiguous, using only one person is justifiable. Regarding the degree of commerciality, we classified the documents as follows:

1. A product is sold on the page
2. Webpage of a company
3. Page of a public authority, government office, administration, etc.
4. Page of a non-governmental organization, association, club, etc.
5. Private page
6. Page of an institution of higher education
7. Other

Classes 1 and 2 were considered as having a commercial intent, while classes 3–7 were considered as non-commercial.

Additionally, we checked the pages for advertisements. We counted the banner ads as well as the text-based ads. (i.e., sponsored links).

---

[1] MSN is the name of Microsoft's online portal. The underlying search engine was Live.com until 2009, from then on Bing. To avoid confusion, we refer to this search engines simply as "Microsoft".

# 5. RESULTS

As stated earlier, we designed a standard retrieval effectiveness test and added a results classification to it. In this section, we describe our results. First, we present the results on the retrieval effectiveness of the search engines under investigation, and then we focus on the differences between commercial and non-commercial results.

## 5.1 Retrieval Effectiveness

First, we compared the macro precision of the search engines—i.e., the number of queries that an individual search engine is able to answer better than its competitors (measured by the overall precision). We found that when considering the result sets of 10 results per engine, Google performed best, with 31 queries answered best, followed by Yahoo, with 13 queries answered best, and finally, Microsoft, with only 9 queries answered best (Fig. 1). While this shows that Google overall performs best, one can also clearly see that the performance of the individual search engines heavily depends on the query. None of the engines is able to answer all queries best, or at least the vast majority of them.

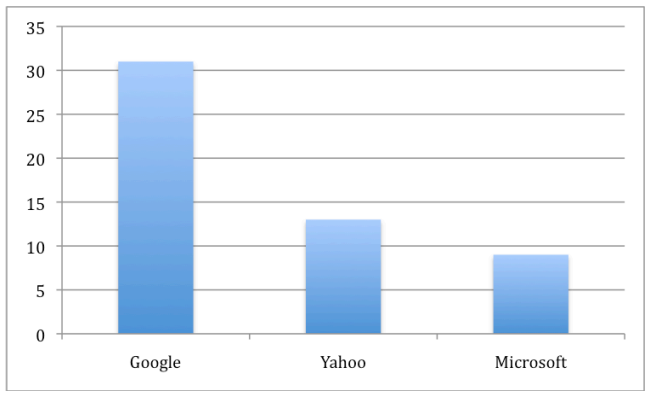

**Figure 1. Macro precision (all search engines; all results).**

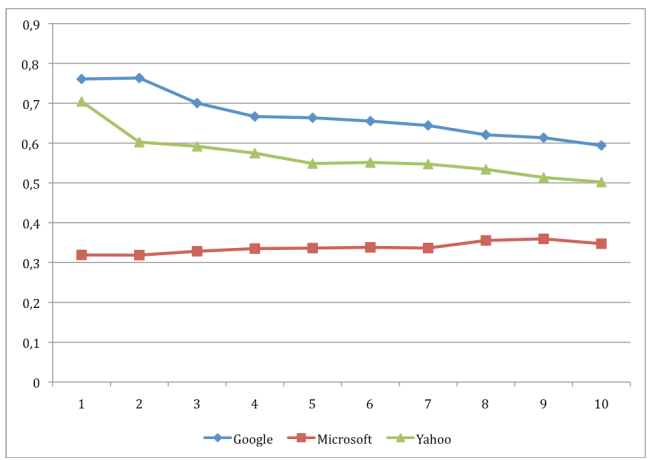

**Figure 2. Precision graph (all search engines; all results).**

Next, we measured the mean average precision of the results according to results position. Figure 2 shows the number of relevant results at position x for the corresponding top x positions. For example, when considering the top three results, precision was calculated for the first three positions *as a whole* (cf. [11]).

One can clearly see that Google outperforms its competitors in terms of mean average precision. This holds true for all results positions. As with macro precision, Yahoo follows in the second rank, while Microsoft comes last.

## 5.2 Ratio of Commercial Results

The classification of all results considered (whether relevant or not) shows that all search engines deliver a comparable number of commercial results (Figure 2). Most notably, web pages provided by a company come in first with between 70% (Yahoo) and 77% (Google) of all results, while pages directly selling products account for only a minority of results. Also, web pages provided by non-commercial institutions or private persons account for only a low ratio of results. Therefore, in the following analysis, we will merge product pages and company web pages into the commercial category, while the remaining categories will be merged into the non-commercial category.

The results show that while Google provides the highest ratio of commercial results, it also provides the results that were judged best by the jurors. This indicates that the commerciality of the results may not have an effect on the relevance judgments.

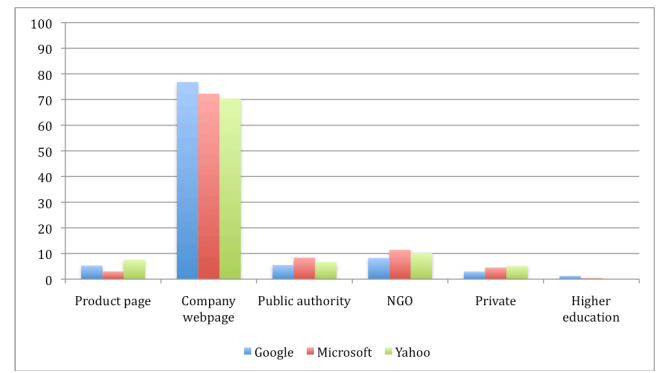

**Figure 3. Ratio of commercial results.**

When looking at the individual results positions, we find that while the ratio of commercial results varies for the individual positions, all engines provide the lowest ratio of commercial results on the first results position (Fig. 4).

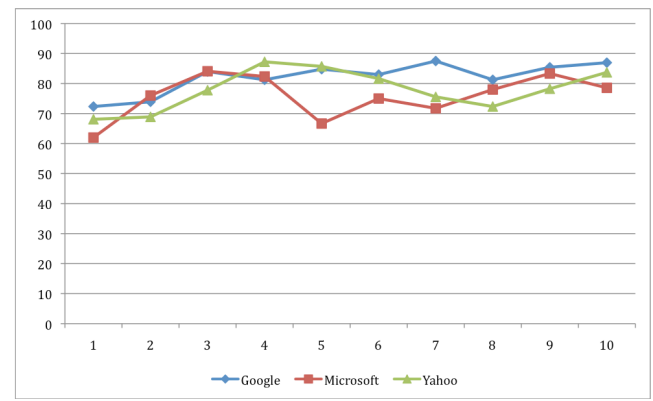

**Figure 4. Ratio of commercial results, according to results position.**

## 5.3 Relevance of Commercial vs. Non-commercial Results

In the preceding sections, we discussed the relevance and the commerciality of the search results separately. In the following sections, we will focus on whether commercial results are judged more or less relevant than non-commercial results, and whether advertisements on the results have an influence on relevance judgments.

Table 2 shows that the ratio of results judged to be relevant does not differ significantly between commercial and non-commercial results. This holds true for all search engines considered.

**Table 2. Ratio of relevant results; commercial vs. non-commercial results.**

| Search engine | Commerciality | Number of results considered[2] | Percentage of relevant results |
|---|---|---|---|
| Google | all results | 442 | 59.4 |
| Google | commercial | 367 | 59.4 |
| Google | non-commercial | 75 | 58.7 |
| Microsoft | all results | 431 | 34.7 |
| Microsoft | commercial | 327 | 34.6 |
| Microsoft | non-commercial | 104 | 36.5 |
| Yahoo | all results | 422 | 50.2 |
| Yahoo | commercial | 333 | 51.1 |
| Yahoo | non-commercial | 89 | 51.7 |

## 5.4 Relevance of Results with Text Ads

Next, we measured the number of results carrying text-based advertising. Note that we considered text ads presented on the results themselves, not the ads on the search engine results pages.

Table 3 gives an overview on the ratio of results with text ads in the top 10 results of the individual search engines. We found that Google provided a significantly higher number of such results than the other search engines did (23.7 percent vs. 12.3 and 17.7 percent, respectively).

| Search engine | Number of results considered | Number of results with text-based ads | Percentage of results with text-based ads |
|---|---|---|---|
| Google | 451 | 107 | 23.7 |
| Microsoft | 444 | 55 | 12.3 |
| Yahoo | 418 | 74 | 17.7 |

**Table 3. Ratio of results with text-based ads**

---

[2] Only results with relevance judgments as well as commerciality judgments are considered. Results not classified as either commercial or non-commercial ("other") are omitted.

Again, we asked whether results with text-based ads are considered more or less relevant than results without such advertising. Comparing the relevance judgments differentiated accordingly, we found that with Google and Yahoo, there is no difference in the relevance judgments. However, the results for Microsoft differ clearly. While 40% of the pages with text ads are judged relevant, only 33.9% of the pages *without* text ads are seen as relevant (Table 3).

**Table 3. Ratio of relevant results; pages with text-based ads vs. pages without text-based ads.**

| Search engine | Text-based ads | Number of results considered | Percentage of relevant results |
|---|---|---|---|
| Google | text ads | 107 | 59.8 |
| Google | no text ads | 344 | 59.3 |
| Microsoft | text ads | 55 | 40.0 |
| Microsoft | no text ads | 389 | 33.9 |
| Yahoo | text ads | 74 | 52.7 |
| Yahoo | no text ads | 344 | 52.0 |

Further analysis concerning the number of ads on the individual results would be interesting, but due to the relatively low number of occurrences, we refrained from this form of analysis.

## 5.5 Relevance of Results with Banner Ads

The ratio of results with banner ads is given in Table 4. Contrary to the ratio of results with text-based ads, there are no differences between the engines with banner ads.

**Table 4. Ratio of relevant results; pages with banner ads.**

| Search engine | Number of results considered | Number of results with banner ads | Percentage of results with banner ads |
|---|---|---|---|
| Google | 451 | 98 | 21.7 |
| Microsoft | 444 | 102 | 23.0 |
| Yahoo | 431 | 95 | 22.0 |

While the differences in relevance judgments are not significant with Google and Microsoft, the results for Yahoo show that pages with banner ads are judged to be more relevant than pages without such advertising (Table 5).

**Table 5. Ratio of relevant results; pages with banner ads vs. pages without banner ads.**

| Search engine | Advertising | Number of results considered | Ratio of relevant results |
|---|---|---|---|
| Google | banner ads | 98 | 58.2 |
| Google | no banners | 353 | 59.8 |
| Microsoft | banner ads | 102 | 33.3 |
| Microsoft | no banners | 342 | 35.1 |
| Yahoo | banner ads | 95 | 55.8 |
| Yahoo | no banners | 336 | 49.1 |

## 5.6 Relevance of Results with Advertising in General

When considering all results carrying advertising (whether banner ads, textual ads, or both—Table 6), we found that the ratio of relevant results at Google and Microsoft corresponds to the ratio of relevant results when looking at all results. With Yahoo, the ratio of relevant results is slightly higher for results with ads than for the complete results set (53.7% vs. 50.2%).

**Table 6. Ratio of relevant results; pages with any form of advertising**

| Search engine | Number of results considered | Ratio of relevant results |
|---|---|---|
| Google | 177 | 58.2 |
| Microsoft | 139 | 33.8 |
| Yahoo | 151 | 53.6 |

## 6. CONCLUSION

From our results, we have seen that in terms of retrieval effectiveness, Google performs best of the search engines compared (RQ1). There is only a slight variation in the ratio of commercial results from search engine to search engine (RQ2). The results classified as "commercial" are comparably relevant to the results classified as "non-commercial" (RQ3). Finally, textual or banner ads do not have a negative effect on the perceived relevance of the corresponding result (RQ4).

We found that Google shows significantly more commercial results than the other search engines under investigation. However, this does not affect the relevance of the results. Further research is needed to find whether Google deliberately boosts these results. As Google is not only the most-used search engine, but also the largest provider of text-based ads, it would be understandable if Google would do so. When deciding between two equally relevant results, it would make sense to prefer the result that also carries text-based advertising from the site's own advertising system. However, further research (using a much larger data set) is needed to verify this assumption.

Our research shows that the top 10 search engine results are highly commercial, even though the queries used did not necessarily have a commercial intent. Again, it would be interesting to classify a larger amount of results for this purpose. Studies collecting large numbers of results from commercial search engines do show that a large ratio of results come from the .com domain [18].

Our research is limited in that we only used a relatively low number of queries and there may have been a bias due to the choice of topics. Furthermore, all these queries were in German. Also, it would be interesting if our results held true if jurors were asked to judge the relevance of the results on a scale instead of giving binary judgments.

While our study surely has its limitations, we think that it provides a good starting point for further investigations of the relevance of certain types of results. Combining retrieval effectiveness tests with results classification (not necessarily a classification of commercial intent) seems a promising way to help improve search results, whether in commercial web search engines or any other information retrieval system.